\documentclass[twocolumn]{aastex631}

\usepackage{hyperref}

\begin{document}

\title{Proto-NUX: A prototype telescope for ground-based near-ultraviolet observations}

\author[0009-0009-7461-1633]{Rasjied Sloot}
\affiliation{Anton Pannekoek Institute for Astronomy, University of Amsterdam, Science Park 904, 1098 XH Amsterdam, The Netherlands \\}
\correspondingauthor{Rasjied Sloot}
\email{m.r.sloot@uva.nl}

\author[0000-0002-3516-2152]{Rudy Wijnands}
\affiliation{Anton Pannekoek Institute for Astronomy, University of Amsterdam, Science Park 904, 1098 XH Amsterdam, The Netherlands \\}

\author[0000-0002-6636-921X]{Steven Bloemen}
\affiliation{Department of Astrophysics/IMAPP, Radboud University, P.O. Box 9010, 6500 GL, Nijmegen, The Netherlands\\}

\author{Rik ter Horst}
\affiliation{NOVA, Netherlands Research School for Astronomy,
    P.O. Box 9513, NL-2300 RA Leiden, The Netherlands}
    
\author{Hans Ellermeijer}
\affiliation{Technology Center, University of Amsterdam,  Science park 904, 1098 XH Amsterdam, The Netherlands}

\author{Alexander Hoogerbrug}
\affiliation{Anton Pannekoek Institute for Astronomy, University of Amsterdam, Science Park 904, 1098 XH Amsterdam, The Netherlands \\}

\begin{abstract}
The Near-UV-eXplorer (NUX) is a proposed ground-based, wide-field telescope array with a field of view of $\sim$70 square degrees, designed to operate over the 300--350~nm wavelength range and to achieve a target sensitivity of 20~mag in 150~seconds ($5\sigma$). Its main scientific objective is the detection and characterization of hot, rapidly evolving transients in the near-UV (NUV). Proto-NUX is a pathfinder instrument for NUX, based on an off-the-shelf  36~cm Celestron RASA wide-field astrograph that has been modified to enhance throughput and image quality in the targeted NUV band. The main objectives of Proto-NUX are: (1) to quantify the NUV sensitivity of the prototype and assess the feasibility of the full NUX facility; and 2) to characterize atmospheric extinction in the NUV, including its temporal variability and its dependence on zenith angle. Using three filter configurations, we aim to measure the wavelength dependence of the atmospheric extinction and to disentangle the contributions from Rayleigh scattering (dominating at wavelengths $>$325~nm) and molecular ozone-dominated absorption (dominating $<$315~nm). On-site testing is scheduled for 2026 at the Pic du Midi Observatory (France, 2877~m altitude) in order to evaluate on-sky performance under high-altitude observing conditions.

\end{abstract}


\section{Introduction}

Over the past decades, time-domain astronomy has developed into a major discipline in astrophysics, with numerous facilities systematically surveying the variable sky across nearly the full electromagnetic spectrum, from radio to gamma-rays 
\citep[for reviews, see, e.g.,][]{2016mks..confE..13F,2019PASP..131a8002B,2019Galax...7...28R}. These surveys have revealed a rich population of transient phenomena associated with a broad range of astrophysical sources, ranging from stellar flares and accretion-driven outbursts of compact objects and black holes (from stellar-mass to supermassive ones) to energetic cosmic explosions.  The latter provide direct observational access to extreme physical conditions, relativistic outflows, and rapid  energy-release processes. Despite the large number of dedicated transient search programs, no wide-field, high-cadence survey is currently operating in the ultraviolet (UV) regime.

The absence of past and currently active UV transient survey facilities is primarily due to  strong atmospheric attenuation of UV radiation, which severely limits ground-based observations, and the substantial financial and technical requirements associated with deploying wide-field UV instrumentation in space, typically at costs of several hundred million Euros. At present, only one dedicated wide-field mission concept, Ultrasat\footnote{www.weizmann.ac.il/ultrasat/}, is under active development \cite[][]{2014AJ....147...79S,2024ApJ...964...74S}.  The absence of systematic UV surveys is particularly limiting, as the UV band provides critical constraints on the physics of many high-energy transients, several of which exhibit pronounced UV emission and may peak at UV wavelengths during the initial stages of their evolution \citep[see the in-depth discussions in][]{2014AJ....147...79S,2021arXiv211115608K,2024ApJ...964...74S}

Although ground-based observations can, in principle, be performed down to wavelengths of $\sim$300 nm, most non-UV-optimized observatories and instruments 
generally have poor performance below 400 nm. This limitation arises primarily from the low quantum efficiency of conventional detectors and suboptimal throughput of optical components at near-UV (NUV) wavelengths. However, recent advances of UV-sensitive detector technology have enabled quantum efficiencies approaching 50\% in the NUV regime, thereby making scientifically competitive instrumentations in this band increasingly feasible \citep[see for example][]{SPIE_Guy_Japan}. To further improve the quality of NUV Observations, optical designs and (mirror) coatings must be specifically optimized for observations at NUV wavelengths. 

The Near-Ultraviolet eXplorer (NUX) is a proposed, specialized wide-field telescope array consisting of four telescopes designed to survey the sky from the ground for NUV transients \citep{NUX_paper}. It is conceived as a transient discovery facility with a $\sim$70 square degree field of view and operates in the 300--350~nm wavelength range (the NUX-band). The primary scientific objective is to improve our understanding of the physical processes that power fast (hours-days), hot transients, such as the electromagnetic counterparts of gravitational wave events,  gamma-ray bursts, and
shock-breakout and shock-cooling emission of supernovae. These science goals are closely related to those motivating the Ultrasat mission \cite[][]{2014AJ....147...79S,2024ApJ...964...74S}; however, NUX probes a distinct and redder NUV window (300-350~nm rather than 230-290~nm) and operates from the ground. NUX therefore constitutes an independent and complementary facility in the NUV regime. In the full NUX configuration, the design goal is to reach an AB magnitude of 20 in a 150-second exposure in the NUX-band, enabling up to ten pointings per night for each observed field in surveying mode \cite[for details see][]{NUX_paper}.

\textbf{In the current NUX design, four Celestron RASA (Rowe Ackermann Schmidt Astrograph) telescopes are envisioned. Several current wide-field survey facilities make use of RASA or similar fast optical designs, including the ATLAS–Teide telescope \citep{ATLAS_teide}, the ARGUS telescope array \citep{argus}, and the Large Array Survey Telescope \citep{LAST}. These instruments have demonstrated successful operation in high-cadence, wide-field survey regimes, confirming the feasibility of such optical configurations for time-domain astronomy. This provides confidence that a comparable design approach, as adopted for NUX, is well suited for efficient survey performance in the NUV.}

In order for NUX to achieve these science goals, the combination of high sensitivity (which increases the achievable cadence) and a large field of view is required.  With such a facility, these transient events can be found at a high rate and be systematically studied. However, two main uncertainties regarding the feasibility of NUX remain: the technical realization of NUX and the attainable sensitivity of the instrument, which also depends on the atmospheric conditions of the observing site (see Section~\ref{sec:atmosphere}). To address these uncertainties, we have constructed a prototype telescope referred to as proto-NUX, consisting of a single telescope with optical characteristics similar to those of the full NUX design (see Tab. \ref{tab:telescopeproperties}). To allow for broadband optical multiwavelength observations an additional Complementary Optical eXplorer telescope is mounted parallel to proto-NUX (Section \ref{section:COX}). To evaluate the feasibility of the NUX concept, we will conduct test observations with proto-NUX and the accompanied optical telescope from high altitude locations. The scientific objectives and observing strategies of these test campaigns are described in Section~\ref{sec:objectives}.

\section{Atmospheric extinction} \label{sec:atmosphere}

\begin{figure}
    \centering
    \includegraphics[width=0.5\textwidth]{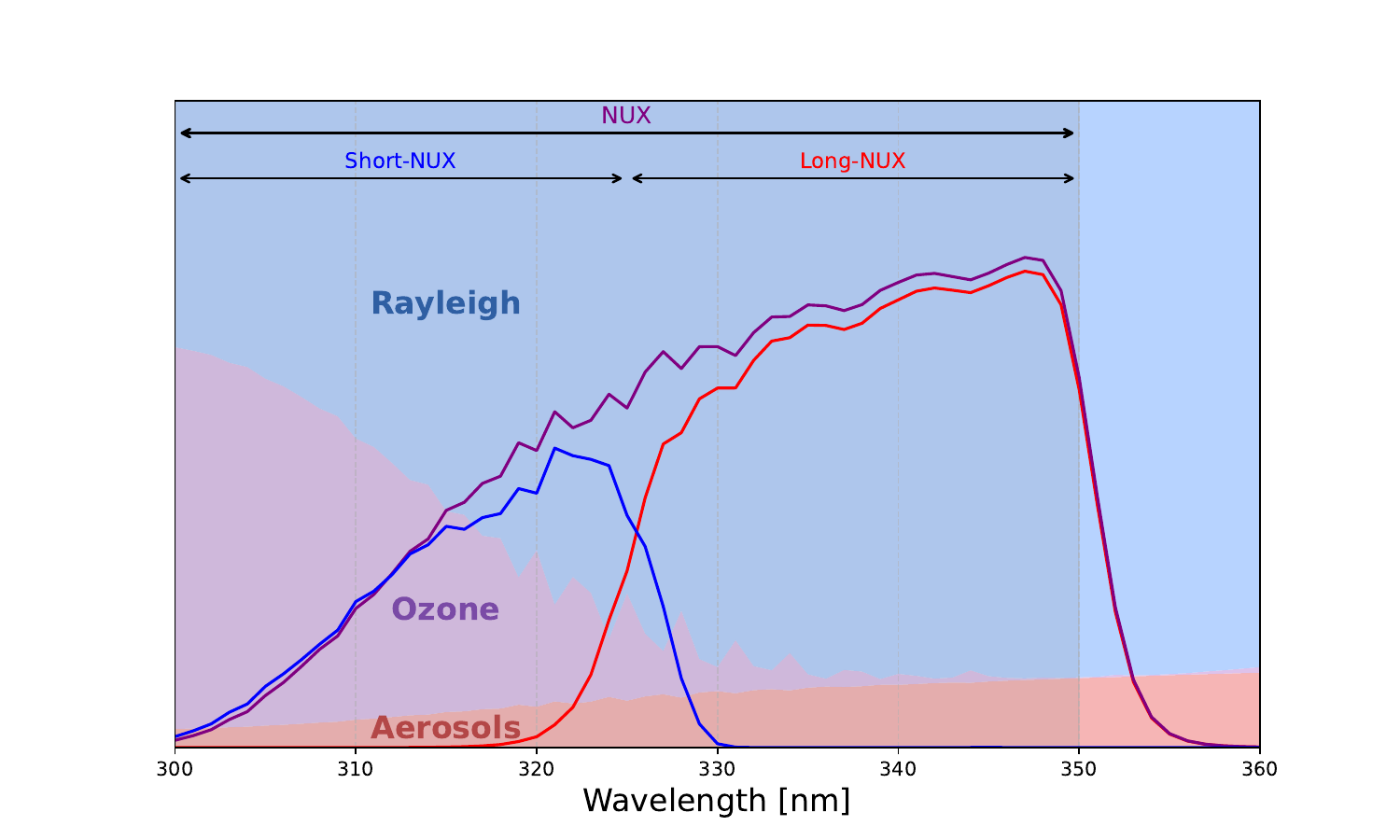}
    \caption{Decomposition of atmospheric attenuation in the NUV into its individual extinction components, calculated with the Cerro Paranal Advanced Sky Model for the La Silla site at 1.2 airmass. Shaded regions indicate the relative contribution of each component to the total attenuation. Vertical lines denote the bandpasses of the adopted NUX filters (see Section~\ref{section:filters}).}
    \label{fig:NUV_atmosphere}
\end{figure}

\begin{table*}[h]
    \centering
    \caption{Comparison between NUX, Proto-NUX, and the Complementary Optical eXplorer. Note that NUX consists of 4 (proto-NUX type) telescopes with slightly overlapping field of views. Therefore the surface area on sky is about 4 times larger. The read noise for the QHY2020 camera is stated for the proposed manufacturere gain setting of 6. The dark current is stated for an operation temperature of -10 degrees.}
    \begin{tabular}{lccc}
        \hline
        & NUX (Concept) & Proto-NUX & Complementary Optical eXplorer  \\
        \hline
        Telescope type & Celestron RASA 36 cm & Celestron RASA 36 cm& Askar 140 w/0.8x reducer \\
        Aperture & 360 mm & 360 mm & 140 mm \\
        Focal Length (effective) & 790 mm & 790 mm & 720 mm \\
        Focal Ratio & f/2.22 & f/2.22 & f/5.6 \\
        Camera & QHY6060 Pro & QHY2020 w/UV-window & ZWO ASI183MM Pro \\
        Sensor (CMOS) & Gpixel GSENSE6060 & Gpixel Gense2020 BSI & Sony IMX183CLK-J  \\
        Sensor size ($\mathrm{mm}^2$) & 61.3x61.3 & 13.3x13.3 & 13.2x8.8 \\
        Pixel size (micron) & 10.0 & 6.5 & 2.4 \\
        Total pixel area & 6134x6134 & 2048x2048 & 5496x3672 \\
        Plate scale ("/pixel) & 2.6 & 1.7 & 0.6 \\
        Field of View ($\mathrm{deg}^2$) & \text{4.45°} x \text{4.45°} & \text{0.97°} x \text{0.97°}  &  \text{0.96°} x \text{0.64°}  \\
        Surface on sky  & 19.8 $\mathrm{deg}^2$ & 0.96 $\mathrm{deg}^2$  & 0.61 $\mathrm{deg}^2$ \\ 
        Filters & t.b.d. & NUX, Short-NUX, Long-NUX & u,g,r,i,z,y,L \\
        Read Noise (e-)  & $>4.7$ & $~3.4$ & 2.3 \\
        Dark Current (e-/sec/pix) & N.A. & 0.07 & 0.004 \\
        \hline
    \end{tabular}
    \label{tab:telescopeproperties}
\end{table*}

Atmospheric extinction in the NUV (i.e., in the NUX band of 300–350 nm) remains poorly constrained, primarily due to the limited number of observational studies conducted at these wavelengths. To investigate the dominant components contributing to atmospheric extinction, we employed the Cerro Paranal Advanced Sky Model\footnote{Web version available at: www.eso.org/sci/observing.html} \citep{SkyCalc1,SkyCalc2} to generate representative atmospheric extinction profiles. Since La Silla Observatory in Chile is the targeted site for the full NUX facility \citep[see][]{NUX_paper}, this sky model was adopted as a proxy for the expected atmospheric conditions. The resulting extinction curves indicate that, in the NUX band at an airmass of 1.2 airmass, extinction is dominated by Rayleigh scattering at wavelengths longer than $\sim$325~nm and by ozone absorption at shorter wavelengths (in particular below $\sim$315 nm), with an additional contribution of about 5\% from aerosols via Mie scattering throughout the NUX band (see Fig. \ref{fig:NUV_atmosphere}).

Combining atmospheric transmission with modeled instrument throughput yields a predicted $5\sigma$ limiting magnitude of $\sim$20 AB in 2.5 minute exposure at zenith under dark conditions \citep{NUX_paper}. The simulations further indicate that sensitivity losses remain below $\sim$0.5 mag up to zenith angles of $\sim$45$\degr$, and below $\sim$1 mag up to $\sim$60$\degr$. These estimates assume identical zenith-angle scaling for Rayleigh scattering and ozone absorption, which introduces systematic uncertainties. Studies of solar UV radiation indicate that ozone absorption exhibits a stronger zenith-angle dependence than Rayleigh scattering, implying wavelength-dependent variations in extinction across the NUX band \citep{Hawaii_UV_zon}.  A comparable effect is expected for nighttime observations.

The temporal variability at night in NUV extinction is poorly characterized, particularly in the ozone layer. Existing u-band extinction monitoring at La Silla shows significant seasonal, nightly, and intranight variability driven mainly by aerosol and meteorological effects \citep{Atm_la_Silla}. At shorter wavelengths (under $\sim 325$ nm), extinction is expected to be increasingly sensitive to (tropospheric) ozone concentration variations, which exhibit diurnal, seasonal, and inter-night fluctuations. Currently, it is unclear if these variations are strictly correlated with the extinction variability in the u-band.

To quantify these effects, proto-NUX will conduct observations in the full 300–350 nm band as well as in two sub-bands (300–325 nm and 325–350 nm; hereafter referred to as the short-NUX and long-NUX band, respectively; see Fig. \ref{fig:NUV_atmosphere}). These measurements will enable an empirical characterization of wavelength-dependent extinction, its scaling with zenith angle, and its temporal variability, thereby providing the calibration framework required for accurate NUX photometry (see Section~\ref{sec:sensitivity}).

\section{the Proto-NUX Project: a prototype ground-based near-UV telescope}

\subsection{Telescope, optics and detector}

\begin{figure}
    \centering
    \includegraphics[width=1\linewidth]{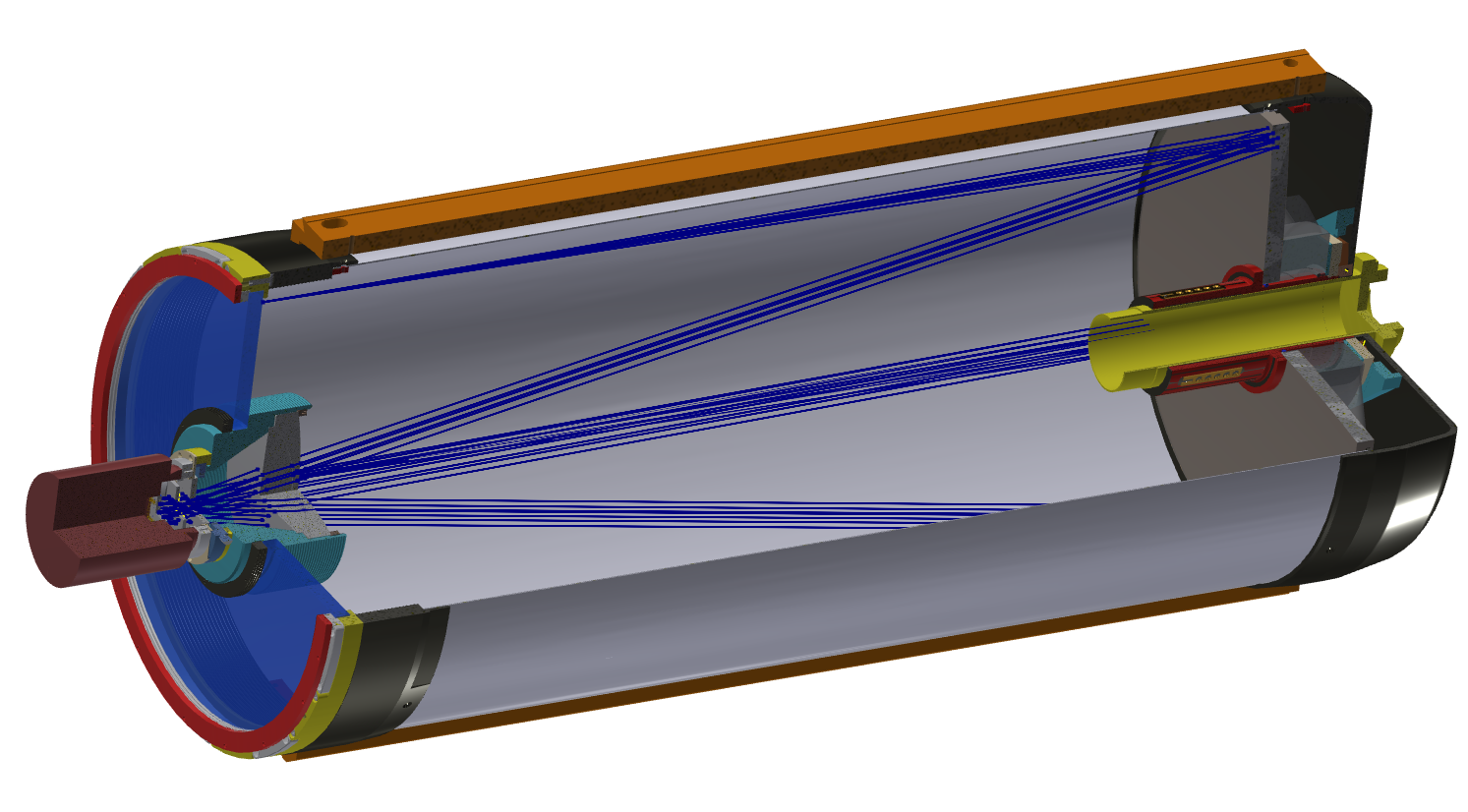}
    \caption{3D drawing of the proto-NUX telescope. The optical tube and focusing mechanism are the original components of an off-the-shelf 36 cm Celestron RASA telescope. The front section of the telescope (left side of the image) has been completely redesigned and newly manufactured, including the Schmidt-corrector (blue), the corrector holder (black/yellow), and the lens assembly (teal). The CMOS camera (brown) is attached to the lens assembly via a tilt plate and a filter slider. A close up view is shown in Fig. \ref{fig:lens_assembly}.}
    \label{fig:PNUX_geheel}
\end{figure}

\begin{figure}
    \centering
    \includegraphics[width=1\linewidth]{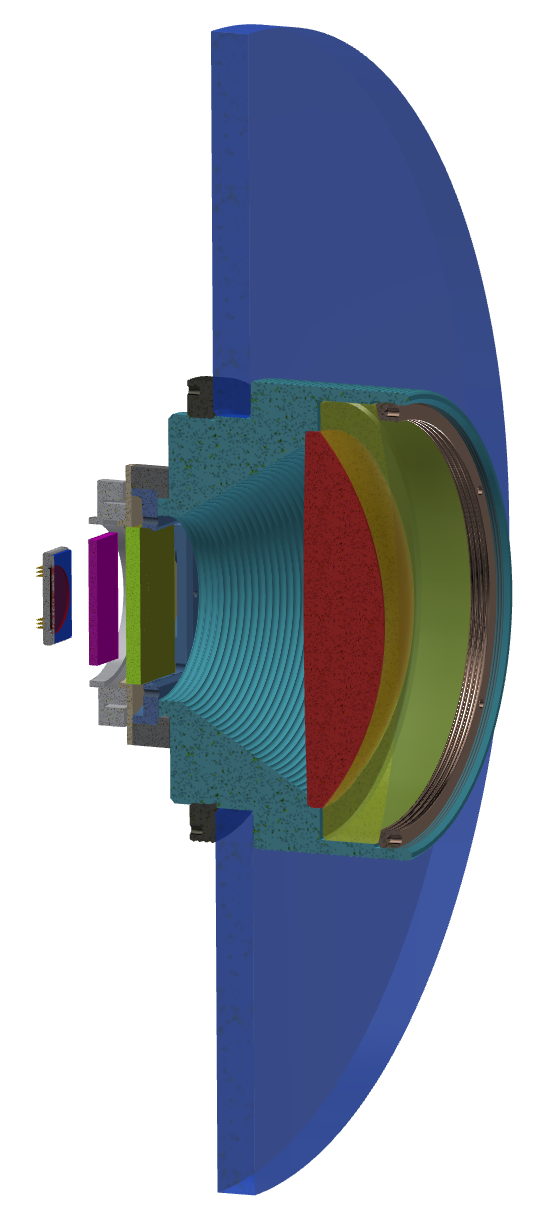}
    \caption{Close-up 3D view of the redesigned lens assembly. The Schmidt corrector (blue) supports the aluminum lens assembly housing (teal), which is secured by a retaining ring (black). The two lenses (red/yellow) are held in place by a threaded retaining ring (gold). A filter slider (green) and the CMOS detector are attached to the housing; the camera sensor and its protective UV-transparent window (purple) are shown.}
    \label{fig:lens_assembly}
\end{figure}

\begin{figure}
    \centering
    \includegraphics[width=1\linewidth]{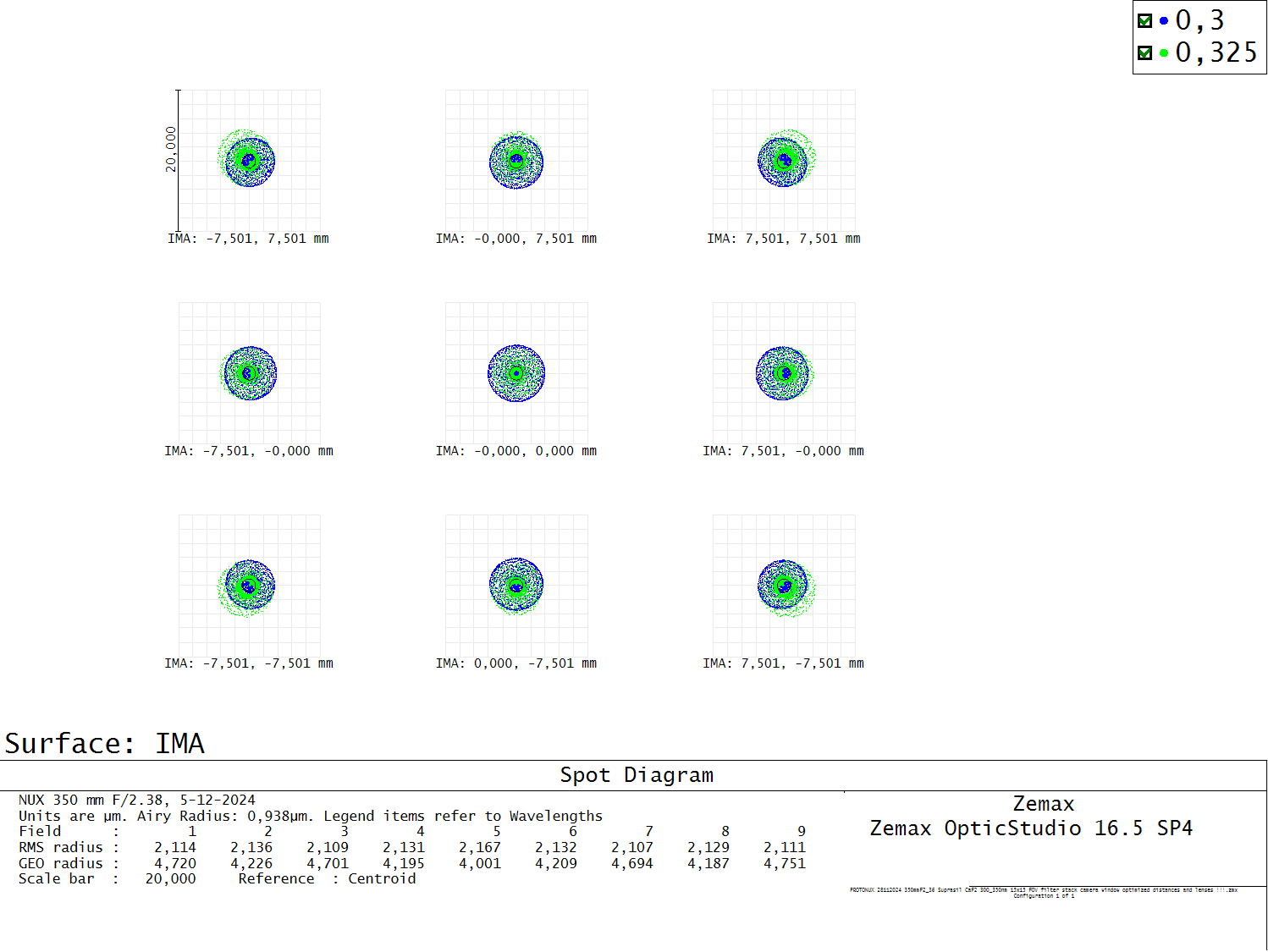}
    \includegraphics[width=1\linewidth]{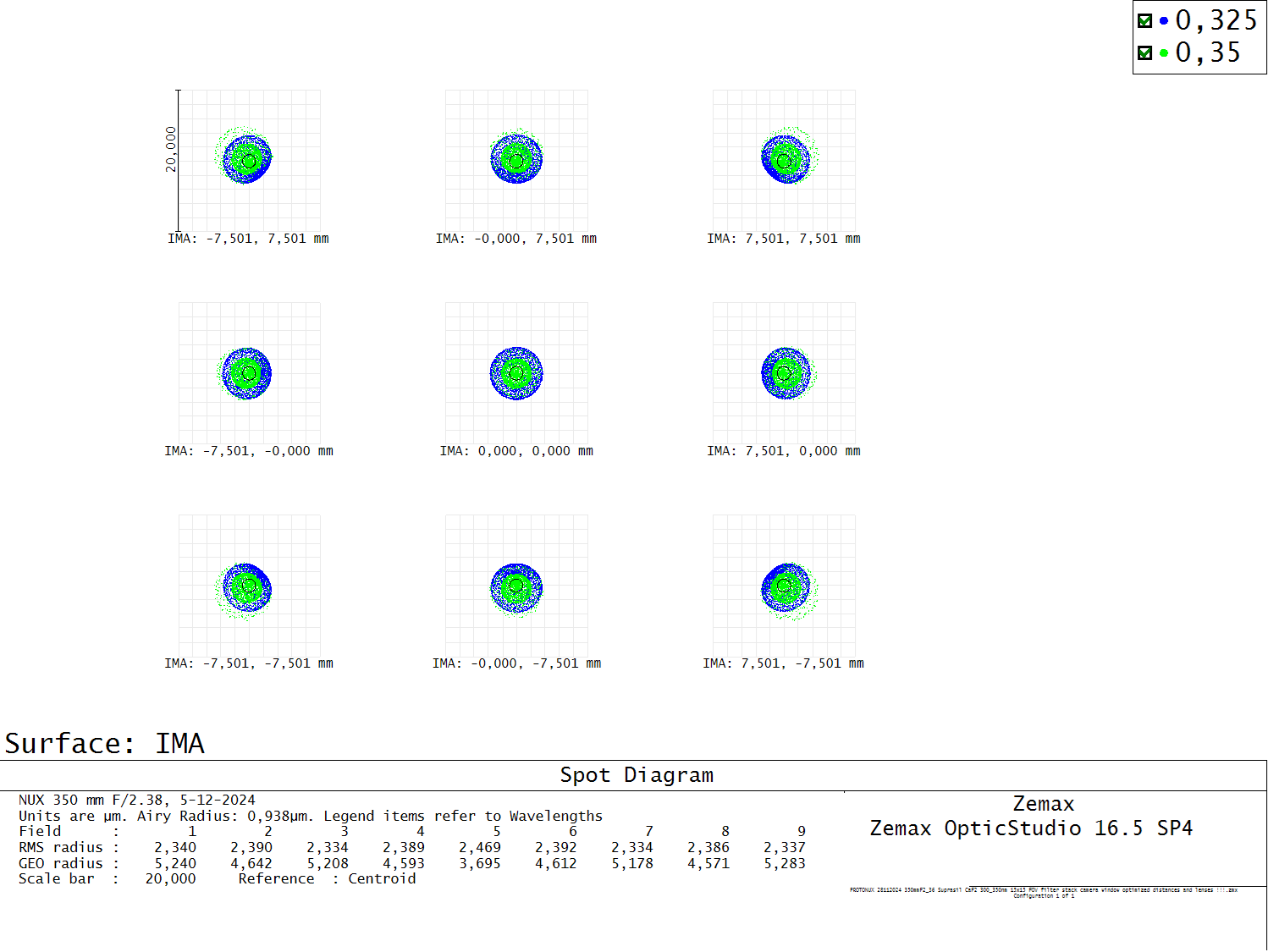}
    \caption{\textbf{Spot diagrams of the proto-NUX optical design at different field positions across the field of view, roughly corresponding to the size of the Gpixel GSENSE2020 BSI sensor. The diagrams are shown for multiple wavelengths (top figure: 300nm and 325nm, bottom figure: 325nm and 350nm). The black circle represents the Airy disk (0.938 micron), and the squares are about three times the pixel size (6.5 micron) of the CMOS detector. The enclosed energy (EE) is above 80\% within a diameter of 6.5 micron across the field.}}
    \label{fig:spot}
\end{figure}

Proto-NUX consists of a single telescope (in contrast to the full NUX facility, which comprises four identical telescopes; \citealt{NUX_paper}) and is based on a 36 cm Celestron RASA design. After procurement of the standard commercial telescope, the optical configuration was redesigned to maximize sensitivity and image quality in the NUX band (see Fig. \ref{fig:spot}). The original Schmidt corrector plate was not transparent in the NUV, necessitating a redesign of this optical component. 

A new Schmidt corrector was designed and fabricated from fused silica, implemented as a plane-parallel optical window that compensates for the spherical aberration of the spherical primary mirror, consistent with the classical Schmidt and original RASA configurations. The required aspheric deformation is applied to both surfaces (the same for both surfaces) such that, in combination with the new lens group (see below), optimal correction of chromatic aberration, spherical aberration, coma, and astigmatism is achieved, with the system specifically optimized for performance within the NUX spectral range. A detailed description of the process of fabrication of the corrector is provided in \cite{NUX_paper}. 

As a consequence of the redesign, the corrector had to be positioned at a larger distance from the primary mirror than in the original configuration, necessitating a mechanical modification of the original Schmidt holder ring at the front of the optical tube assembly. The holder ring retains the original Celestron dew-heater system to prevent condensation from forming on the Schmidt corrector. 

In addition, the original primary mirror contained Starbright XLT optical coatings\footnote{www.celestron.com/pages/starbright-xlt-optical-coatings} which includes an aluminum reflective layer and multiple layers of mirror coating, which exhibits relatively low reflectivity in the NUX band \citep[estimated at 15\%-50\%,][]{Coating_reflection}. To improve performance at 300-350 m, the primary mirror was removed from the optical tube assembly and recoated. The factory coating was not stripped; instead a new layer of bare aluminum with a thickness of $\sim$120~nm was deposited directly on top of the existing coating. Subsequently, a multilayer UV-enhanced overcoat of $\sim$130~nm was applied on top of the bare aluminum layer. Both coatings were applied via vapor-deposition in a vacuum chamber by Sumipro\footnote{www.sumipro.nl}, a company based in Almelo, The Netherlands. 

Furthermore, the original four-element extra-low dispersion glass lens group located at the prime focus was replaced by a custom-designed two-element lens assembly (see Fig. \ref{fig:lens_assembly}), consisting of CaF$_2$\footnote{Produced by VY Optoelectronics Co., Ltd. www.vyoptics.com} and fused silica\footnote{Produced by Synoptix sinoptix.eu/}. The mechanical housing for the new lens group was redesigned and fabricated from black-anodized aluminum at the Technology Center of the Faculty of Science at the University of Amsterdam. 

Attached to the prime-focus lens group assembly, the remainder of the optical train consists of several off-the-shelf components obtained from Baader Planetarium\footnote{www.baader-planetarium.com/} in Germany and are an M56 tilter, a UFC filter base with filter sliders accommodating 50$\times$50~mm filters (see Section \ref{section:filters} for details), and an M54a camera adapter. The detector is a QYH2020UV BSI camera\footnote{Supplied by QHYCCD www.qhyccd.com} (see Tab. \ref{tab:telescopeproperties} for specifications) equipped with a quartz entrance window to maximize quantum efficiency in the NUX-band. The average quantum efficiency in the 300-350 nm range is approximately 50\%.

\subsubsection{Filter configurations and total system throughput} \label{section:filters}

\begin{table*}[h]
    \centering
    \caption{Filter configurations used in proto-NUX. The individual filters are combined into three filter configurations: NUX-band, Short-NUX, and Long-NUX. The peak transmission includes the effects of all optical components and the atmospheric transmission (1.2 airmass, La Silla). The relative throughput is measured with respect to the NUX-band configuration. The red leak values of the three different filter configurations are estimated assuming a flat spectrum and blackbody spectra of various temperatures.}
    \begin{tabular}{lllccc}
        \hline
        Filters & Type & Transmission range (nm) &NUX-band & Short-NUX & Long-NUX \\
        \hline
        ZHS0350T & Shortpass & 300-350 & X & & X \\
        ZUV0325T & Shortpass & 250-325 &  & X &\\
        ZUL0325T & Longpass & $>$325&  & & X\\
        ZRR0340T & Red Rejection & 290-380 & X & X & X\\
        \hline
        
        Peak transmission (system) & & & 18.9\%  & 11.9\% & 18.4\%\\
                \hline

        Relative throughput \\(compared to NUX-band) & & & 100\%  & 29\% & 69\%\\
                \hline

        Red Leak &Flat spectrum & & 0.42\% & 0.76\% & 0.83\% \\
        & Blackbody spectrum & &  &  &  \\
          &- 3 000 K  & & 22\% & 47\% & 30\% \\
          &- 6 000 K  & & 0.20\% & 0.41\% & 0.36\% \\
          &- 10 000 K  & & 0.031\% & 0.054\% & 0.062\% \\
          &- 40 000 K   & & 0.005\% & 0.008\% & 0.012\% \\
          &- 100 000 K  & & 0.004\% & 0.006\% & 0.009\% \\
        \hline
  \end{tabular}
    \label{tab:telescopesetups}
\end{table*}

\begin{figure}
    \centering
    \includegraphics[width=1\linewidth]{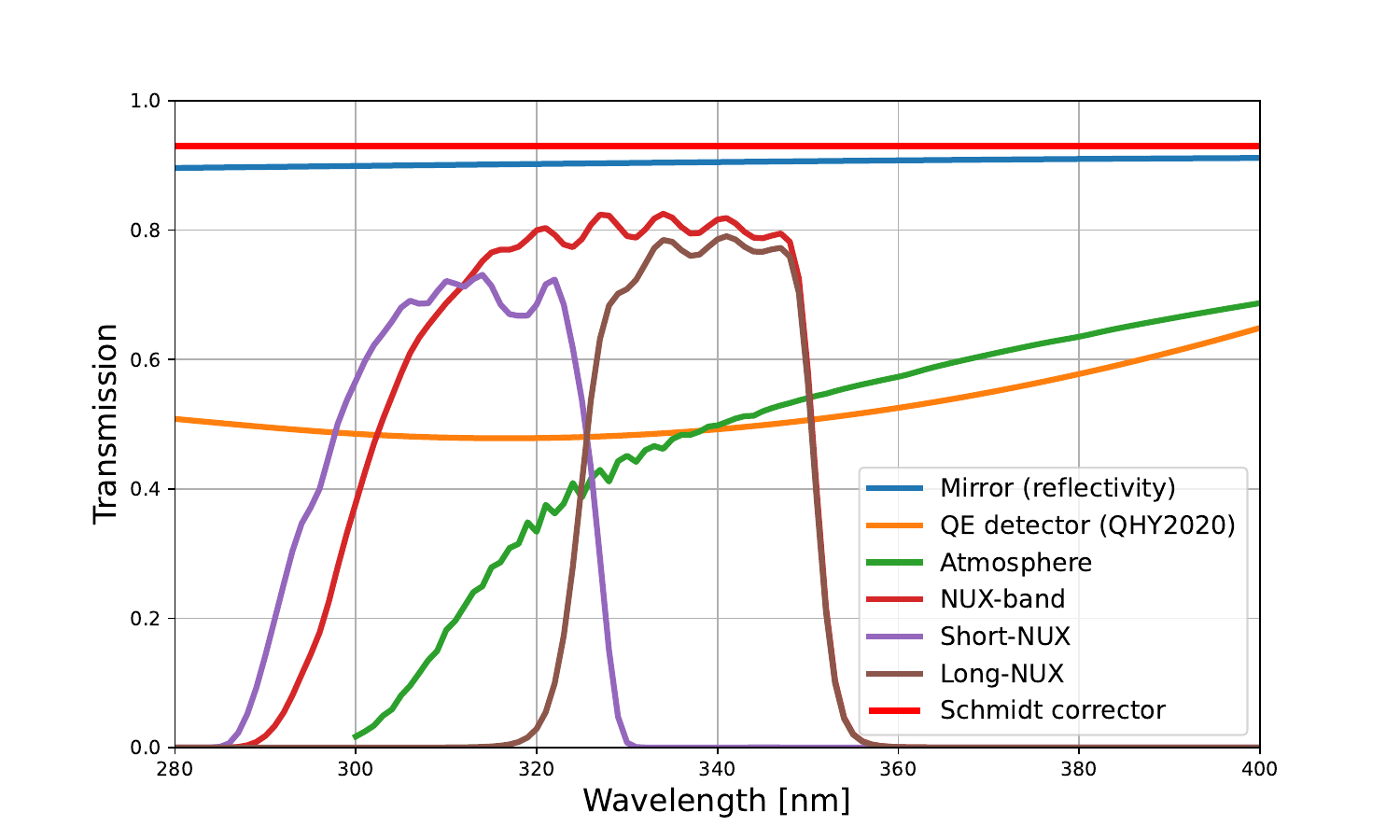}
    \caption{Transmission curves of the individual components contributing to the proto-NUX system throughput. The figure shows the transmission of the three NUX filter configurations, the atmospheric transmission (1.2 airmass, la Silla), the mirror coating reflectivity, the quantum efficiency of the CMOS camera, and the transmission of the Schmidt corrector. See the inset for the color coding of the different components.}
    \label{fig:transmissionfilters}
\end{figure}

\begin{figure}
    \centering
    \includegraphics[width=1\linewidth]{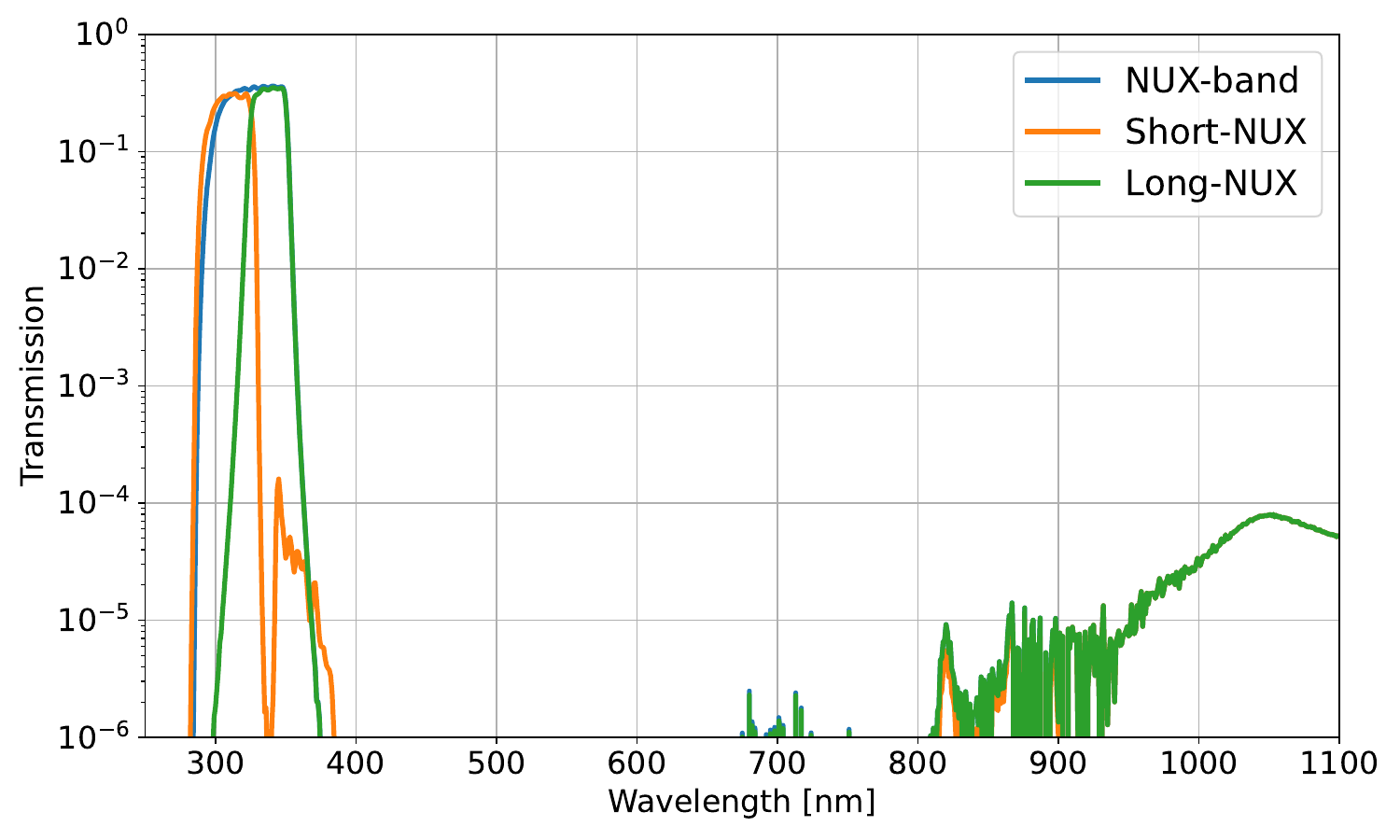}
    \caption{Transmission curves of all three NUX filter configurations showing the 'red leak' at longer wavelengths. The total contribution of the red leak is $<0.9\%$ of the total system throughput, measured between 200 and 1100 nm. See the inset for the color coding of the different filter configurations.}
    \label{fig:redleak}
\end{figure}

Proto-NUX employs off-the-shelf filters from ASAHI Spectra\footnote{www.asahi-spectra.com}. To obtain the desired bandpasses and to suppress the intrinsic red leak commonly present in such filters, multiple filters are stacked in each configuration.  Observations in the NUX-band configuration constitute the default operating mode for proto-NUX and provided a bandpass of 300--350~nm. This bandpass is achieved by combining the ZHS0350 filter with a red blocking filter (ZRR0340) to suppress the intrinsic red leak of the primary bandpass filter\footnote{For the original transmission curve of the ZHS0350 filter, we refer to www.asahi-spectra.com/opticalfilters/detail.asp?key=ZHS0350 .}. \textbf{The red-leak contributions of the filter configurations have been calculated for a range of black-body temperatures spectra by integrating the transmission profiles. The results are summarized in Table~\ref{tab:telescopesetups} and show that for spectra above $5000K$ the red-leak filter transmission will be $<1\%$ (e.g. less than 1 percent of the photons passing through the filter configuration are outside the filter bandpass). When the temperature of the observed target decreases below $5000K$, the red leak becomes significantly worse (see Tab. \ref{tab:telescopesetups}). This makes proto-NUX particularly suitable for studying hot ($>5000K$) transients at high (better than 0.01 mag) photometric accuracy. }

To distinguish between different scattering and absorption processes in the Earth's atmosphere, which may exhibit uncorrelated variability in NUV transmission (see Section \ref{sec:atmosphere} for details), two additional filter configurations are implemented that subdivide the NUX-band. The short-NUX configuration covers a bandpass of 300–325~nm, where atmospheric attenuation is primarily governed by ozone absorption, while the long-NUX configuration covers 325–350~nm, where attenuation is dominated by Rayleigh scattering. All filters are housed in a 50$\times$50mm filter slider by Baader Planetarium, which is operated manually. All filter configurations used are summarized in Table \ref{tab:telescopesetups} and corresponding system throughput curves are shown in Figure~\ref{fig:transmissionfilters}.

We note that the filter concept for proto-NUX was briefly introduced in \citet[][]{NUX_paper}. However, due to mechanical constraints of the filter slider mechanism, 50$\times$50 mm filter versions were adopted instead of the previously considered 25 mm diameter filters. In addition, for cost considerations, the long-NUX band was not implemented as a custom-manufactured filter \cite[as initially envisaged in][]{NUX_paper}, but instead constructed from a combination of three off-the-shelf filters, as listed in Table~\ref{tab:telescopesetups}.

\subsubsection{Sensitivity Estimate}
\label{section:performance}

Proto-NUX is designed to closely follow the specifications of NUX. However, due to differences in the detector choice, its expected performance is not identical to that reported by \citet{NUX_paper}, which quotes a limiting AB magnitude of 20 for a 150s exposure. A dedicated sensitivity estimate is therefore required for the Proto-NUX configuration.
The full-system sensitivity has been calculated by combining the measured transmission curves of the optical system (Fig.~\ref{fig:PNUX_geheel}) with detector and telescope parameters provided by the manufacturer. The total system throughput includes contributions from the Schmidt corrector, primary mirror, relay optics, and detector quantum efficiency. Atmospheric transmission is modeled using the La Silla SKYCALC tool, ensuring representative observing conditions for potential NUX sites.

The sky background is adopted from U-band measurements at La Silla presented by \citet{La_silla_tes_giraud}. For the near-ultraviolet regime, a value of $22\mathrm{mag/arcsec^{-2}}$ is assumed, representative of a clear, moonless night. This value is used consistently throughout the sensitivity calculation.

The limiting magnitude is derived using the standard CCD signal-to-noise formalism \citep[Chapter~4]{Howell_CCD}. The signal-to-noise ratio is computed by including contributions from the source signal, sky background, detector dark current, and read noise. Photometry is performed within a circular aperture with a radius of $1.5\times$ the full width at half maximum (FWHM) of the stellar point-spread function, which optimizes the balance between enclosed signal and background noise (see Tab. \ref{tab:telescopeproperties}).

\begin{figure}
    \centering
    \includegraphics[width=1\linewidth]{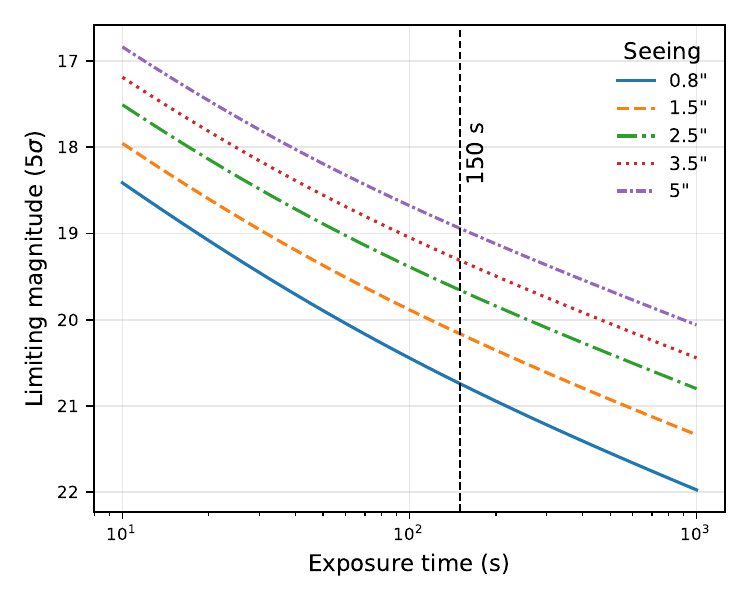}
    \caption{Limiting magnitude as a function of exposure time for the NUX bandpass of the proto-NUX telescope. Curves are shown for different seeing conditions. The limiting magnitude is defined at a detection threshold of $5\sigma$. Assumptions for noise contributions are described in section \ref{section:performance}. A vertical dashed line indicates an exposure time of $150\mathrm{s}$ which is the default exposure time proposed for the full NUX. Sensitivity for the 'Short-NUX' and 'Long-NUX' bandpasses (see Tab. \ref{tab:telescopesetups} and Fig. \ref{fig:transmissionfilters}) are 0.9 and 0.4 AB magnitude worse, respectively.}

    \label{fig:sensitivity_curves_seeing}
\end{figure}

The resulting sensitivity ($5\sigma$ limiting magnitude) as a function of exposure time and seeing conditions is presented in Fig.~\ref{fig:sensitivity_curves_seeing}. These calculations provide a realistic estimate of the on-sky performance of Proto-NUX under typical observing conditions at sites with properties similar to La Silla. We note that uncertainties in system throughput and sky background introduce a systematic uncertainty in the derived sensitivities. In addition, seasonal variations in atmospheric extinction have been measured at La Silla in the U-band at the level of approximately 0.2--0.3 mag \citep{la_silla_extinction_rufener}. In the NUV we expect these variation to be larger, but by how much larger needs to be determined by long term monitoring using proto-NUX. 

\begin{figure}
    \centering
    \includegraphics[width=1\linewidth]{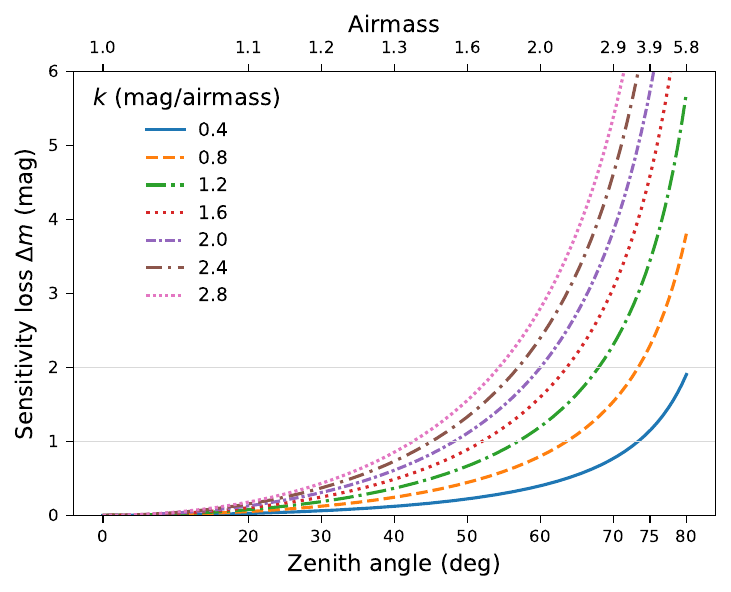}
    \caption{Sensitivity loss as a function of zenith angle for different atmospheric extinction coefficients $k$ (in mag/airmass). The curves are computed using an $X=1/cos(z)$ airmass approximation. The sensitivity loss is defined as $\Delta m = k (X - 1)$, where $X$ is the airmass. The horizontal reference lines indicate sensitivity losses of 1 and 2 magnitudes. The top axis shows the corresponding airmass values.}

    \label{fig:Sensitivity_loss_airmass}
\end{figure}

To quantify the effect of atmospheric extinction on the limiting sensitivity, we additionally consider the dependence of signal attenuation on zenith angle. The extinction-induced magnitude loss is given by $\Delta m = k (X - 1)$, where $k$ is the wavelength-dependent extinction coefficient and $X$ is the airmass computed using a $X=1/cos(z)$ approximation where $z$ is the zenith angle. Figure~\ref{fig:Sensitivity_loss_airmass} illustrates this dependence for representative extinction values. The relation highlights that observations at higher airmass can incur significant sensitivity losses, reaching several magnitudes under high-extinction conditions, especially when observering using the short-NUX filter configuration. Typical La Silla conditions of $k \sim 0.4$--$0.6$ in the optical, increasing toward the near-UV, a 1 magnitude impact becomes particularly relevant at zenith angles above $\sim 40^\circ$.

\subsection{Complementary Optical eXplorer}
\label{section:COX}

\begin{figure}
    \centering
    \includegraphics[width=1\linewidth]{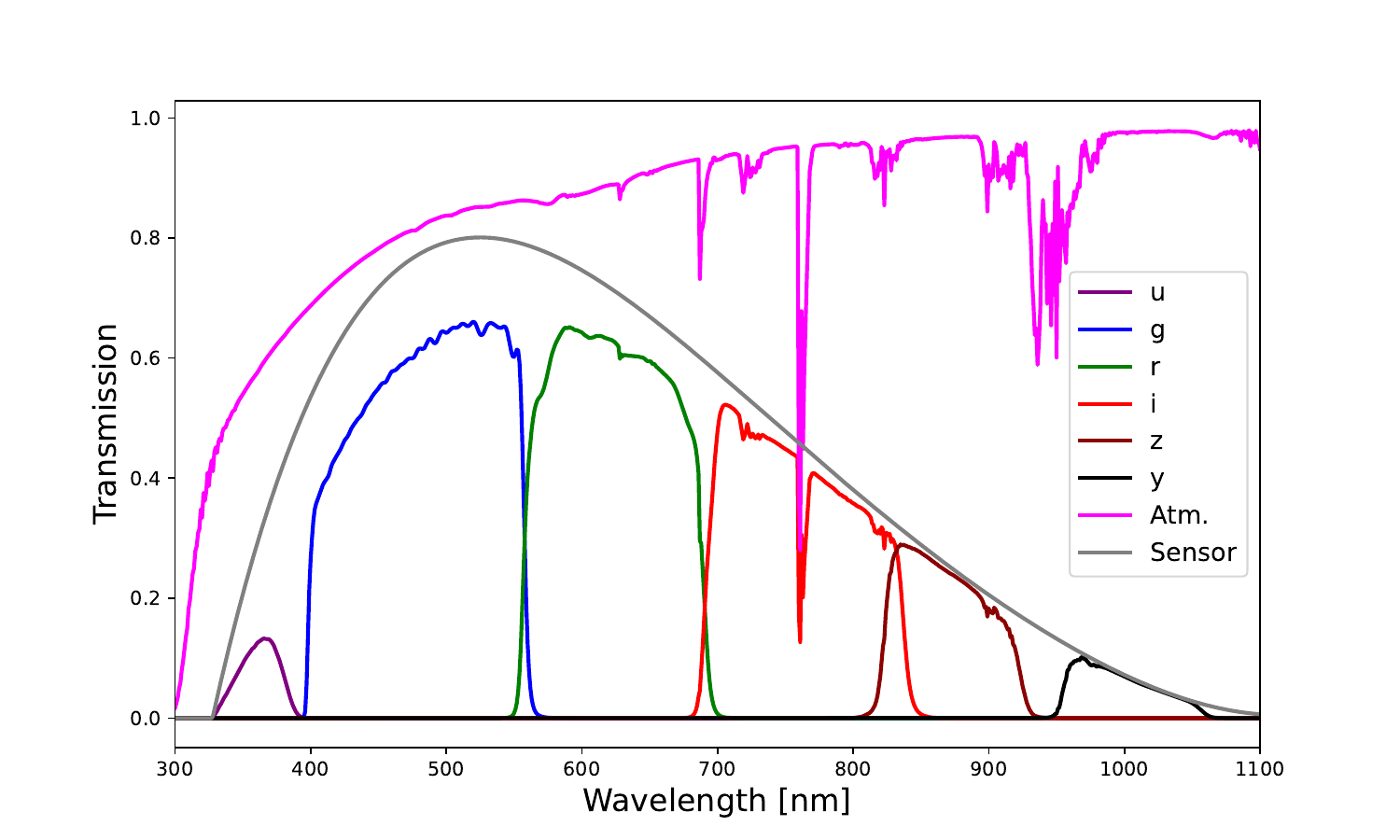}
    \caption{Transmission curves of all available filters in Complementary Optical eXplorer. The transmission of the filters have been convoluted with the atmospheric transmission at la Silla Observatory (1.2 airmass) and with the QE of the ASI183MM pro camera (including protective sensor glass). Transmission through the telescope optics have not been taken into account since the exact type of glass and coatings used are unknown. Therefore, the transmission in the u-band could be even less than depicted in the figure, particularly below 380 nm.}
    \label{fig:placeholder}
\end{figure}

The primary objective of the full NUX facility is the discovery and identification of very blue transients. To distinguish genuine blue transients from intrinsically red, but very bright, transients that exhibit a strong blue tail in their spectral energy distribution, a complementary optical telescope operating at longer wavelengths is envisaged to function in parallel with NUX. In addition, such an instrument provides an independent means to quantify any residual red leak in the NUX system.

\begin{figure}
    \centering
    \includegraphics[width=1\linewidth]{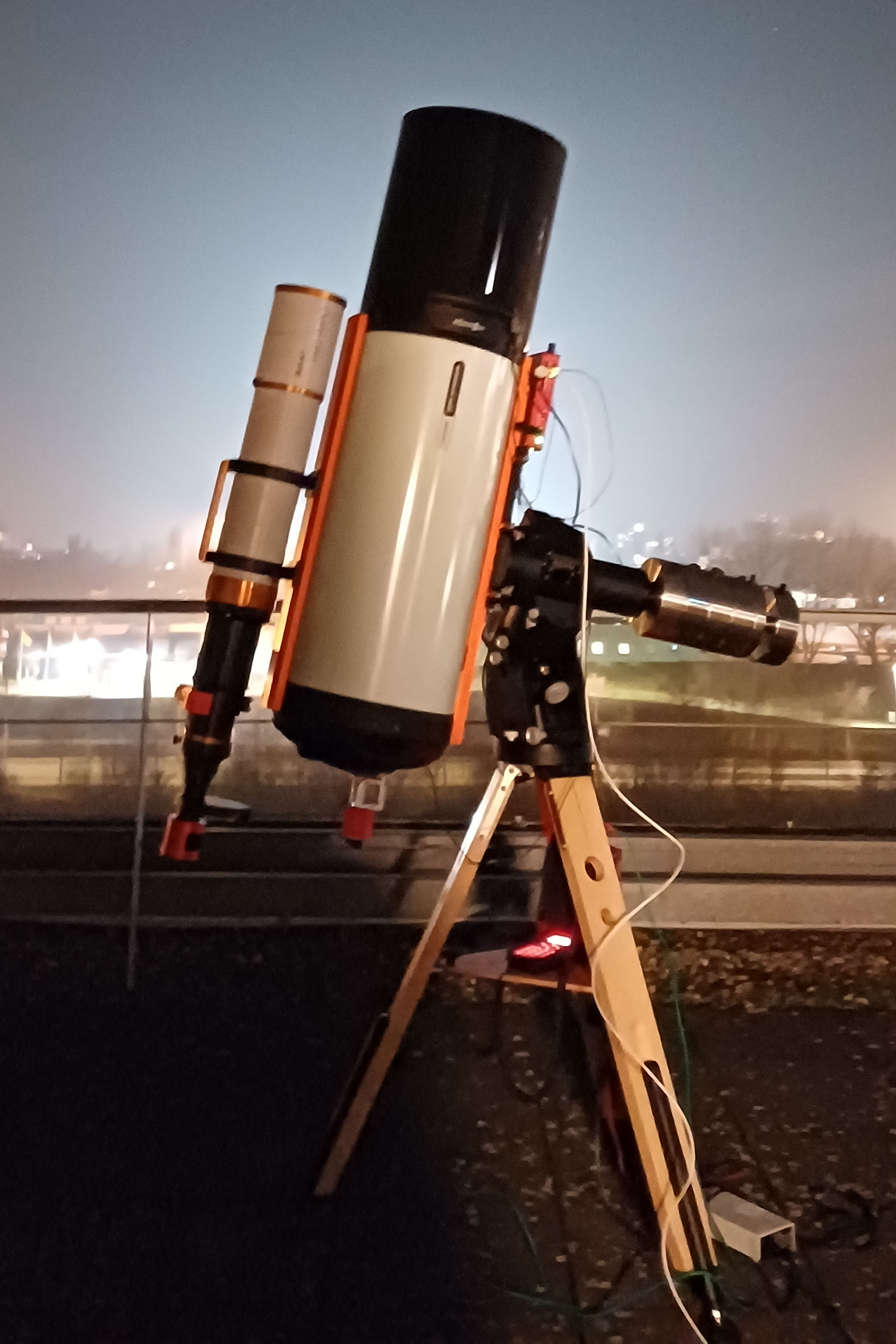}
    \caption{The proto-NUX setup during a test night on 4 March 2026 at the Anton Pannekoek Observatory, University of Amsterdam. Proto-NUX is equipped with a dew shield (black). The Complementary Optical eXplorer is piggyback-mounted on proto-NUX. All hardware is controlled by the Eagle PC (red), mounted near the dew shield.}
    \label{fig:pnux_opgesteld}
\end{figure}

To test the simultaneous operation of this dual-instrument configuration, we acquired a commercial ASKAR 140 mm APO f/7.0 triplet refractor equipped with a 0.8$\times$ reducer/flattener, an eight-slot filter wheel, and a ZWO ASI183MM Pro camera. The filter wheel is populated with six Baader Planetarium SDSS filters (u, g, r, i, z, y) and a clear L filter covering the wavelength range 380–1050 nm.  Focus adjustments during filter changes are performed using an electronic focuser (ZWO EAF). The telescope will be mounted in parallel with proto-NUX (see Fig.~\ref{fig:pnux_opgesteld}) and provides a comparable field of view (see Table~\ref{tab:telescopeproperties}).

In addition for the proto-NUX telescope, continuous u-band monitoring with the optical companion provides an important reference for atmospheric characterization. Since proto-NUX cannot observe the NUX, short-NUX, and long-NUX configurations simultaneously, and (manual) filter changes may be limited during a given night, the u-band offers a stable comparison channel. Because the transmission in the u-band and the long-NUX band are both primarily affected by Rayleigh scattering, their correlated behavior can be used to determine whether extinction variations in the ozone-dominated short-NUX band tracks the Rayleigh driven changes or instead exhibit independent variability (see Section~\ref{sec:atmosphere}).

Beyond validating simultaneous operation, this configuration enables characterization of residual red-leak contamination in proto-NUX and facilitates contemporaneous multi-band observations of candidate science targets (see Section~\ref{sec:sciencetargets}).

\subsection{Non-optical equipment}

To facilitate testing at multiple observation sites, the proto-NUX system is designed as a portable and transportable setup that can be easily shipped and deployed. The complete proto-NUX setup is transported in a custom-made flight case to ensure mechanical protection during shipping. The telescope is mounted on a 10Micron GM2000 HPS II COMBI equatorial mount installed on a Geoptik Pegasus tripod. The mount has a payload capacity of 50 kg and is equipped with high-precision absolute encoders, enabling accurate pointing and tracking without the need for an external autoguider. This configuration ensures reliable and stable telescope operation under field conditions.

System control is provided by a PrimaLuceLab Eagle 5 XTM, a Windows-based mini-PC mounted directly on the telescope. This unit manages mount control, camera operations (including power distribution), and the anti-dew heating system. To mitigate the impact of potential power interruptions during remote deployments, dedicated battery packs are available to supply power to the entire system. Image acquisition, telescope pointing, and automated focus are performed using the MaximDL software package\footnote{diffractionlimited.com/product/maxim-dl/}

\section{Proto-NUX Objectives and Observation strategy} \label{sec:objectives}

This section describes the primary objectives of proto-NUX and the observing strategies designed to achieve them. The prototype serves as both a technical demonstration for the full NUX facility and as a platform to quantify the key performance parameters that determine its scientific viability. In particular, we aim to characterize instrumental sensitivity, atmospheric effects in the NUX wavelength regime, photometric stability, and optical performance under realistic observing conditions. The observing strategy has therefore been developed to systematically isolate and measure these factors, while also enabling limited science verification during the commissioning campaigns.

\subsection{Observing runs}

Following the modification and replacement of the original telescope optics, proto-NUX was assembled in January 2026 at the University of Amsterdam. Initial testing was conducted in February and early March 2026 at the Anton Pannekoek Observatory (located sea level; \citealt{APO_paper}). During these commissioning tests, the telescope was collimated using bright stars, and all hardware required for high-elevation deployment was tested and prepared for shipment.

Testing in Amsterdam allowed verification of simultaneous camera control for proto-NUX and its optical companion telescope, execution of automated observing scripts, and validation of operational procedures including polar alignment, mount tracking accuracy, dew prevention, focusing, and guiding performance.
 
To evaluate performance under reduced atmospheric column densities, proto-NUX will be deployed to high-altitude sites. The first campaign will take place at the Pic du Midi Observatory in the French Pyrenees (2877 m altitude). This site was selected for its high elevation, accessible infrastructure, and logistical feasibility from the Netherlands. Potentially, additional tests will be conducted at La Silla Observatory, Chile (2400 m altitude), the preferred long-term site for the full NUX facility \citep[see][]{NUX_paper}, owing to its favorable observing conditions, including a high fraction of clear nights and stable atmospheric performance.

\subsection{Sensitivity} \label{sec:sensitivity}

\paragraph{Objective:} Determine the 5$\sigma$ limiting magnitude as a function of exposure time in the three NUX bands, and assess whether the target sensitivity of AB = 20 in a 150 s exposure in the NUX band can be achieved. 

\paragraph{Observation strategy:} A field near the celestial pole or close to zenith is preferred to minimize differential airmass variations during long integrations. Exposure times ranging from 1 s to 900 s will be used to determine the limiting magnitude as a function of integration time. The selected field should contain multiple blue stars, such as young stellar clusters. In the northern hemisphere, the open cluster NGC 2281 ($\delta=41^\circ$, $m_V=5.4$) is suitable, as it passes near zenith at the beginning of the night.

Instrument sensitivity also depends on atmospheric extinction (zenith-angle dependent) and sky background. These effects are addressed in Sections \ref{sec:atmostransmis} and \ref{sec:skybackground}

\subsubsection{Atmospheric transmission} \label{sec:atmostransmis}

\paragraph{Objective:} Measure extinction coefficients ($k$) in all three NUX bands, expressed in AB magnitudes per unit airmass.

Extinction coefficients have been measured at several observatories at wavelengths $\geq 300$ nm. Typical values at La Palma\footnote{La Palma Technical Note 31, 1985}, La Silla \citep{Atm_la_Silla}, and Paranal \citep{paranal} are approximately 0.5 mag/airmass near 350 nm and exceed 1 mag/airmass at shorter wavelengths. However, these represent average values; extinction varies both within and between nights \citep[e.g.,][]{walraven_photometer}. Seasonal trends and enhanced aerosol content (e.g., dust, volcanic activity) may further modify atmospheric transmission \citep[e.g., see][]{Atm_la_Silla}.

\paragraph{Observation strategy:} To measure $k$, a stellar field containing a sufficient number of bright stars will be observed over a broad airmass range. The goal is to determine extinction coefficients with a precision of 0.01 mag/airmass under stable conditions. A circumpolar field is preferred to allow extended coverage across large airmass variations without rapid altitude changes. If not, the object would set and pass through the `high airmasses' too fast, leaving not enough time for longer (10-15 min) exposures without having too much differential airmass. Objects that culminate at around 3-4 airmass (15$\degr$ altitude) are favorable. For Pic du Midi, suitable targets include NGC 884 or NGC 869, which are circumpolar and span approximately 1.3–5.4 airmass during the night. NGC 2281 provides an alternative option.

The chosen object could be observed for as long as possible over the full range of airmasses while alternating between NUX filters to compare extinction behavior across bands\footnote{Amsterdam test runs demonstrate that filter changes and repointing require less than one minute.}. If necessary to improve precision, a single-filter-per-night strategy will be adopted to maximize exposure statistics. Default exposures will be 150 s, corresponding to the targeted AB = 20 design sensitivity \citep{NUX_paper}. Measurements will be repeated over multiple nights to assess intra- and inter-night variability in $k$.

\subsubsection{Sky background and Moonlight} \label{sec:skybackground}

\paragraph{Objective:} Measure the NUV sky background and quantify its dependence on lunar phase.

The sky background in the NUX bands is a key determinant of instrument sensitivity. Near the detection threshold, read noise, and sky background flux dominate the noise budget. Under dark conditions, primary contributors to the NUV background\footnote{Estimated using ESO SkyCalc: www.eso.org/observing/etc/bin/simu/skycalc} include airglow, scattered starlight, and zodiacal light. When the Moon is above the horizon, scattered moonlight is expected to dominate. During twilight, scattered sunlight becomes the principal source of background.

\paragraph{Observation strategy:} The Moon will be observed in the NUX bands to measure its brightness and characterize scattered light across the sky at varying angular separations. This will enable calibration of sky brightness as a function of lunar phase and geometry. Twilight sky brightness will be measured at zenith during flat-field acquisition. The time interval between sunset and stabilization of the NUV sky background will define the usable observing window, which is expected to extend longer than in redder optical bands.

\subsection{Band selection}

\paragraph{Objective:} Determine the optimal bandpass configuration for the final NUX facility.

NUX is designed to operate between 300 and 350 nm, making the full NUX band the primary test configuration. However, atmospheric attenuation within this band is governed by two dominant processes (Rayleigh scattering and ozone absorption; see Fig.~\ref{fig:NUV_atmosphere}), which may exhibit different temporal variability (see Section~\ref{sec:atmosphere}). This may complicate accurate photometric calibration for the NUX band.

To investigate this, the short-NUX and long-NUX configurations have been implemented, isolating ozone- and Rayleigh-dominated wavelength regimes, respectively. Because most of the transmission within the NUX band originates from the long-NUX region, it is important to evaluate whether inclusion of the short-NUX band provides sufficient scientific gain to justify potential calibration complexity.

The trade-off lies between higher sensitivity (from broader or redder throughput) and potentially reduced photometric stability due to ozone variability.

\paragraph{Observation strategy:}  Extinction coefficients will be measured across multiple nights to quantify intra-night and night-to-night variability in all bands. Simultaneous u-band photometry from the optical companion telescope will provide a Rayleigh-dominated reference channel. Since the u-band has negligible sensitivity within the NUX bandpass, it serves as an independent tracer of atmospheric transparency. By comparing variability patterns between long-NUX (Rayleigh-dominated), short-NUX (ozone-dominated), and u measurements, we will assess whether ozone-driven variations correlate with Rayleigh scattering or exhibit independent behavior.

\subsection{Optical quality}

\paragraph{Objective:} Assess the optical performance of proto-NUX and identify potential limitations relevant for the full NUX design.

The proto-NUX detector has a smaller format and pixel size than the camera planned for the full NUX facility (see Tab.~\ref{tab:telescopeproperties}). While sensitivity can be evaluated, full-field optical performance cannot be completely characterized.  In order to learn whether the whole field will be usable, the point spread function (PSF) will be measured across the detector field to evaluate image quality, field curvature, and residual aberrations. These measurements will be extrapolated to estimate the optical quality of the full field of view. Vignetting across the full image circle cannot be fully tested without a larger sensor.

Multiple stacked filters introduce additional reflective surfaces that may produce ghost images when observing bright stars. To quantify ghosting, long exposures of bright blue stars (e.g., Castor, spectral type A) will be obtained in all filter configurations and without filters. If significant ghosting is detected, optical tilts, anti-reflection coatings on the lenses or design adjustments can be considered for the final NUX implementation.

Seeing typically degrades toward shorter wavelengths and at higher airmass. Optical seeing measurements from the companion telescope will provide reference values. By comparing PSFs in the NUV and optical bands, correlations between NUV and optical seeing can be established. Imaging of extended sources will be used to evaluate spatial resolving performance.

\subsection{Possible science targets on the first observing campaign} \label{sec:sciencetargets}

The primary goal of the Pic du Midi campaign is to achieve the technical objectives outlined above. However, if observing conditions permit additional observations, proto-NUX data may also contribute to astrophysical investigations aligned with the science goals of the full NUX facility \citep{NUX_paper}.

Priority targets include rapidly evolving blue transients such as supernovae, gamma-ray burst afterglows, and accretion-driven outbursts in compact binary systems. Alerts from TNS\footnote{Transient Name Server: www.wis-tns.org/}, GCN\footnote{General Coordinates Network: gcn.nasa.gov/}, and ATel\footnote{The Astronomer's Telegram: www.astronomerstelegram.org/} will be monitored in real time. Targets exhibiting strong blue emission and rapid evolution (timescales of hours to days) will be selected for follow-up. Multi-band light curves will be obtained using both proto-NUX and the optical companion telescope, providing well-sampled spectral energy distributions for comparison with transient models.

In addition to transient follow-up, persistently variable sources will be observed. Variable stars represent an important class for NUV characterization. RR Lyrae stars are particularly suitable targets due to their large ultraviolet amplitudes \citep{Siegel2015}, which exceed typical optical amplitudes. Their pulsational variability produces strong temperature and atmospheric structure changes that are amplified at shorter wavelengths.

We aim obtain time-series observations of RR Lyrae itself using the three proto-NUX bands over a substantial fraction of its pulsation cycle, coordinated with simultaneous multiband measurements in the optical filters spanning u through y. As the prototype of its class, RR Lyrae offers a well-determined ephemeris (see \citealt{2024RAA....24f5021E} and reference therein), large pulsation amplitudes that increase toward ultraviolet wavelengths  \citep[e.g.][]{1982PASP...94..910B}, and favorable visibility from northern observatories, allowing reliable phase-resolved measurements during commissioning. Its radial pulsations produce pronounced variations in effective temperature and atmospheric structure over the $\sim$0.57 d cycle, making it a suitable target both for evaluating sensitivity to short-wavelength variability and for obtaining astrophysically useful constraints. The resulting phase-resolved light curves and spectral energy distributions from the near-UV to the red optical will constrain temperature evolution and atmospheric response through the cycle, provide empirical input for models of $\kappa$-mechanism–driven pulsations in horizontal-branch stars, and demonstrate the capability of the NUX facility for variable-star studies.

\section{Conclusion}

Proto-NUX has been developed as a single-telescope prototype to evaluate the technical feasibility and scientific performance of the full NUX facility. The instrument and its associated hardware have been assembled and commissioned in Amsterdam and are ready for the first high-altitude test campaign at the Pic du Midi Observatory. The results of these tests will determine whether NUX can operate as a ground-based near-UV transient survey facility with the targeted sensitivity and photometric stability, and whether additional site testing (e.g., at La Silla) or design modifications are required prior to construction of the full array. These results will be published in a future paper (Sloot et al., in prep.).

The Pic du Midi commissioning campaign is designed to address several key objectives:

\begin{itemize}

    \item The achievable limiting magnitude as a function of exposure time will be measured in all three NUX bands, directly testing whether the design goal of AB magnitude of 20 in 150 s can be reached under realistic observing conditions. The true sensitivity directly determines the cadence that can be achieved in survey mode. If the targeted depth can be obtained within a few minutes per pointing, NUX will be able to operate at the high cadence required for detecting rapidly evolving transients. However, if substantially longer integrations (e.g., $\sim$10 minutes) are required to reach the same depth, this would significantly reduce survey cadence and may necessitate a redesign of the overall facility concept.

    \item Extinction coefficients and their temporal variability will be quantified to assess the stability of atmospheric transmission in the 300–350 nm regime. The correlation (or lack thereof) between ozone-dominated and Rayleigh-dominated wavelength regions will determine whether a single broad NUX band provides sufficient photometric robustness or whether a narrower long-NUX configuration is preferable for the final facility.

    \item Measurements of sky background and lunar illumination will constrain operational limits and define the effective duty cycle of NUX, directly influencing the survey cadence and the field-selection strategy.

    \item Characterization of optical performance, including PSF stability across the field, ghosting behavior, and seeing dependence, will identify potential limitations in image quality and inform refinements in the optical design, coating strategy, and filter implementation of the full NUX system.

    \item Limited science verification during commissioning will demonstrate the practical capability of proto-NUX to detect and characterize blue transients and variable sources in coordination with complementary optical observations.

\end{itemize}

Depending on the results of these objectives, several paths may be considered for the future development of NUX. These include moving directly toward construction, implementing targeted redesigns of specific subsystems (such as the filter configuration or optical elements), conducting additional site characterization campaigns, or refining the survey strategy. In parallel, we are also exploring the possibility of deploying proto-NUX at a lower-altitude hosting site. This would enable longer-term near-UV monitoring of astrophysical sources and atmospheric transmission properties, further strengthening the foundation for the full NUX facility.

\section{Acknowlegdements}

The proto-NUX project is supported by the NOVA Instrumentation Program in the Netherlands and by the University of Amsterdam (UvA) Stimuleringsbeurs awarded to R.~Wijnands. This research has made use of NASA’s Astrophysics Data System (ADS) Bibliographic Services. Portions of the manuscript text were refined using AI-assisted language tools; all scientific content, analysis, and conclusions are solely the responsibility of the authors.

\section{Data Availability}

The data underlying this article are publicly available and can be accessed via the data reproduction package at DOI: 10.5281/zenodo.19253227 (WILL BE UPDATED AFTER ACCEPTANCE).

\bibliography{PASPsample631}{}
\bibliographystyle{aasjournal}

\end{document}